\documentclass[twocolumn,showpacs,preprintnumbers,amsmath,amssymb]{revtex4}

\usepackage{graphicx}
\usepackage{bm}

\begin{document}

\title{Lateral Shift Makes a Ground-Plane Cloak Detectable}

\author{Baile Zhang}
\author{Tucker Chan}
\author{Bae-Ian Wu}
\email{biwu@mit.edu}

\affiliation{Research Laboratory of Electronics, Massachusetts
Institute of Technology, Cambridge, Massachusetts 02139, USA.}


\begin{abstract}
We examine the effectiveness of the ground-plane invisibility cloak
generated from quasiconformal mapping of electromagnetic space. This
cloak without anisotropy will generally lead to a lateral shift of
the scattered wave, whose value is comparable to the height of the
cloaked object, making the object detectable. This can be explained
by the fact that the corresponding virtual space is thinner and
wider than it should be. Ray tracing on a concrete model shows that
for a bump with a maximum height of 0.2 units to be hidden, the
lateral shift of a ray with 45$^\circ$ incidence is around 0.15
units.
\end{abstract}

\pacs{41.20.Jb, 42.79.-e}

\maketitle

The invisibility cloaks formed by the coordinate transformation
method are rapidly developing from narrowband to
broadband~\cite{pendry,leonhardt,schurig,baile_rainbow,li_carpet,ruopeng,leonhardt_broad,valentine,gabrielli,park}.
For a given coordinate transformation from ``virtual space'' into
``physical space,'' the homogeneous Maxwell equations retain their
form, with only the constitutive parameters changing together with
the field values. This formal invariance gives rise to a powerful
technique for designing optical devices, one of which is the
invisibility cloak. Recently there have been reports on the
applications of quasiconformal mapping to achieve broadband
invisibility by utilizing a ground plane of the perfect electric
conductor (PEC)~\cite{li_carpet,ruopeng,valentine,gabrielli,park}.
This new strategy is based on the assumption that a sufficiently
minimized anisotropy of transformed media can be dropped and
consequently isotropic media in place of anisotropic media can be
used to construct a broadband cloak~\cite{li_carpet}. Subsequent
experiments have implemented this model in both
microwave~\cite{ruopeng} and optical
frequencies~\cite{valentine,gabrielli,park}. Most studies have
considered this kind of ground-plane cloak as being undetectable.
However, the physical consequence as well as the validity of this
quasiconformal mapping technique, i.e., neglecting of the anisotropy
and replacement of anisotropic materials with isotropic materials,
have not been thoroughly discussed in previous studies.

In this Letter we examine the cloaking effectiveness of the
quasiconformal-mapped ground-plane cloak which omits the minimized
anisotropy. We demonstrate that the anisotropy of the transformed
cloak medium is of fundamental importance in preserving both the
phase and energy propagation of electromagnetic waves. Although the
isotropic ground-plane cloak is able to largely preserve the phase
propagation, the energy propagation is subject to a lateral shift
which will render the cloak detectable. We show in a simplified case
that by dropping the minimized anisotropy, although it can be
arbitrarily small, a lateral shift on the order of the height of the
object to be hidden will occur. In the model of the ground-plane
cloak hiding a bump with the maximum height being 0.2 units, we show
by ray tracing that the reflected ray of an incident ray with
$45^\circ$ incidence is laterally shifted on the incident side of
the cloak by a distance of 0.15 units compared to the ideal case.
Our theoretical analysis shows that this ground-plane cloak is
equivalent to a free space thinner and wider  than the ideal case,
causing a lateral shift. This laterally shifting phenomena is
similar to the effect of simply putting a ground plane above the
object to be cloaked. The conclusion of a perceptible lateral shift,
being inconsistent with previous verifications through ray
tracing~\cite{ruopeng}, electromagnetic numerical
simulation~\cite{li_carpet,valentine,gabrielli,park}, and
experiments~\cite{ruopeng,valentine,gabrielli,park}, provides a
balanced view on the performance of a ground-plane cloak.

\begin{figure}
  \begin{centering}
  \includegraphics[width=1\columnwidth,draft=false]{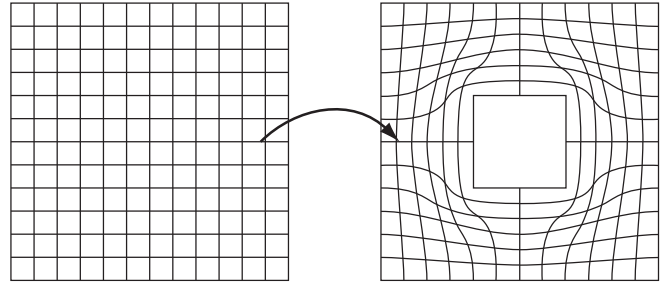}
  \caption{An example of creating an invisibility cloak by coordinate transformation. In general the transformation is not a conformal mapping and the transformed cloak is anisotropic.}
    \label{transforms}
     \end{centering}
\end{figure}

In order to understand the fundamental role of anisotropy in
transformation-based invisibility cloaks, we first recall the
required forms of the constitutive parameters of the transformation
medium. In the transformation medium that is transformed from an
empty space with permittivity $\epsilon_0$ (or permeability
$\mu_0$), the transformed permittivity tensor is
$\overline{\overline{\epsilon}} =\epsilon_0 \overline{\overline{J}}
\cdot \overline{\overline{J}}^T/|J|$ (the permeability
$\overline{\overline{\mu}}$ has similar expression), where
$\overline{\overline{J}}$ is the Jacobian of the transformation. If
we require $\overline{\overline{\epsilon}}$ to be isotropic, the
Jacobian must be a constant times an orthogonal matrix and the
transformation corresponds to a rotation with scaling (here we do
not consider inversion operation). However, for a free standing
cloak~\cite{pendry}, the transformation does not correspond to a
conformal mapping, and therefore orthogonality is not preserved by
the transformation. This has a physical significance and we can
quickly give an interpretation. Consider Fig.~\ref{transforms},
where a transformation is applied to a virtual empty space and a
void is created in the center. In the virtual electromagnetic space,
we can consider the horizontal lines as rays, and correspondingly
the vertical lines as phase fronts. After transformation, the mesh
is distorted. Take the first quadrant as an example. In order to fit
the void in, the horizontal lines in this region are bent clockwise,
while the vertical lines at the same location are bent
counterclockwise. This means the energy propagation is not in the
same direction as the phase propagation, which is not possible in an
isotropic medium. It should be pointed out that a strict conformal
mapping~\cite{leonhardt} does not suffer from this problem, but is
restrictive in the choice of boundary conditions.

\begin{figure}
\begin{centering}
\includegraphics[width=1\columnwidth,draft=false]{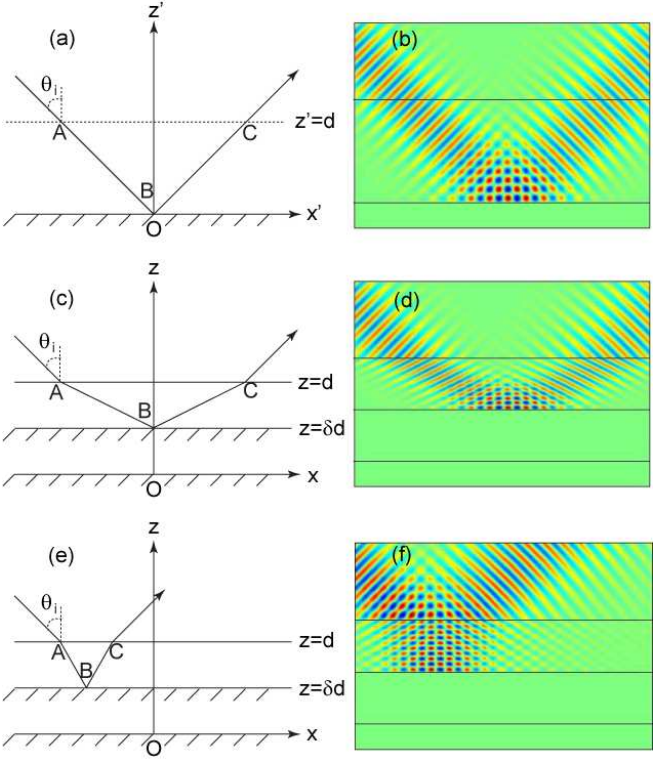}
\caption{\label{planecloak} (Color online) Left column: Ray tracing
diagrams of a ray with incident angle of $\theta_i$. Right column: A
Gaussian beam with incident angle of $\theta_i$. (a),(b): A PEC
ground plane in free space with no cloak. (c),(d): A perfect
anisotropic cloak squeezing space between $z=0$ and $z=d$ into space
between $z=\delta d$ and $z=d$. (e),(f): Replace the anisotropic
cloak with isotropic medium. In the illustration, $\delta = 0.5$ and
$\theta_i = 45^{\circ}$.}
\end{centering}
\end{figure}

For the ground-plane cloak, the anisotropy is also important. Let us
first consider a simplified case, as shown in Fig.~\ref{planecloak}.
Consider a PEC ground plane in free space at the origin, as shown in
Fig.~\ref{planecloak}(a), where the trajectory of a ray with
incident angle $\theta_i=45^{\circ}$ is plotted. Using the
transformation method, it is possible to construct an anisotropic
but homogeneous layer to cloak the entire plane, as shown in
Fig.~\ref{planecloak}(c).  In this transformation, the virtual
electromagnetic space between $z'=0$ and $z'=d$ in
Fig.~\ref{planecloak}(a) is squeezed into the physical space between
$z=\delta d$ and $z=d$ in Fig.~\ref{planecloak}(c) ($\delta=0.5$ in
the illustration). The effect of such a cloak is to have the ground
plane appear to be at the original plane at $z=0$ while in reality
it has been moved to the plane at $z=\delta d$ so that there is room
to hide objects between planes at $z=0$ and $z=\delta d$. The
transformation used to generate this cloak is a one-dimensional
compression: $x = x'$, $y = y'$, and $z =(1-\delta)z'+\delta d$,
where the unprimed variables are the physical space and the primed
variables are the space before transformation. The resulting
relative permittivity and permeability tensors are
$\overline{\overline \epsilon} = \overline{\overline \mu} = \hat
x\hat x \frac{1}{1-\delta}+ \hat y \hat y \frac{1}{1-\delta}+ \hat z
\hat z (1-\delta)$. Note that we can also treat it as an extreme
case of the previous ground-plane cloak model where a curved bump is
changed to be a flat one here~\cite{li_carpet}.

\begin{figure}
\begin{centering}
\includegraphics[width=1\columnwidth,draft=false]{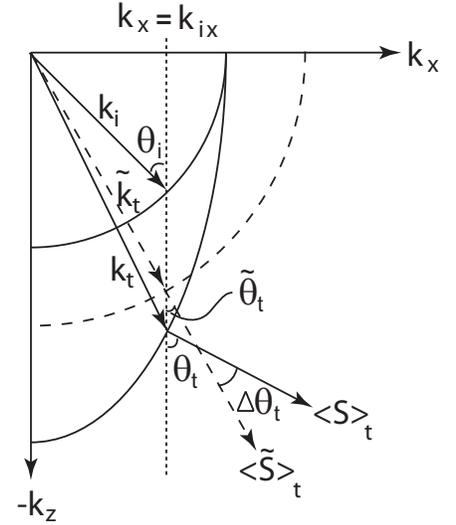}
\caption{\label{phasematching}$k$-vectors used in phase matching
between the incident and cloak media. The solid circle represents
the incident medium and the solid ellipse represents the cloak
medium. The dotted circle represents the isotropic medium that
replaces the cloak medium.}
\end{centering}
\end{figure}

We will first discuss how the cloak in Fig.~\ref{planecloak}(c)
works. For the cloak to be perfect, we require the wave vectors $k_z
= k_z'/(1-\delta)$ and $k_x=k_x'$, such that the total phase after a
round-trip propagation in the cloak layer is the same as when the
cloak is not there, i.e. as in Fig.~\ref{planecloak}(a). Because of
increased $k_z$ after the wave enters the cloak layer, the wave
vector (as opposed to the ray)  is refracted as if the wave has
entered into an optically denser medium with a smaller refracted
angle. Second, we also require the ray to come out at the same point
C in both Fig.~\ref{planecloak}(c) and (a), and therefore the
Poynting vector, or the ray, should follow the path A-B-C in
Fig.~\ref{planecloak}(c). If we take the view of energy or ray
propagation, the light seems to be refracted into an optically less
dense medium with a larger refracted angle.

\begin{figure}
\includegraphics[width=1\columnwidth,draft=false]{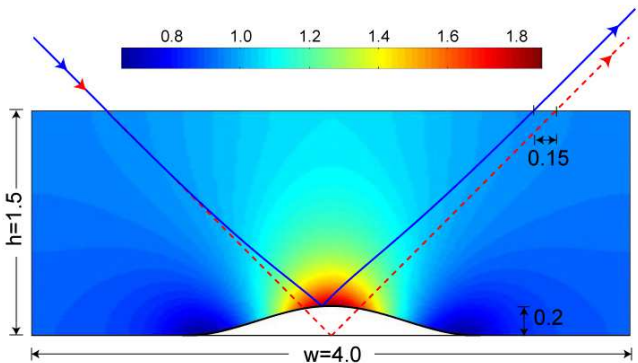}
  \caption{\label{raytracing} (Color online) A ray (blue solid line) incident at $45^{\circ}$ to the normal onto an isotropic ground-plane cloak. The bump has a maximum height of 0.2 units~\cite{li_carpet}. The red dotted line represents the trajectory of an undistorted ray in the absence of the cloak and the bump. The background represents $n^2$. Notice the distorted ray has been shifted on top of the cloak toward the incident point by 0.15 units.}
\end{figure}

This discrepancy between the refracted angles required respectively
by the phase and energy propagation is solved by the anisotropy of
the cloak. In the $k$-surface diagram (Fig.~\ref{phasematching}),
the incident medium, free space, gives a circle of radius $\omega /
\text{c}$ and the cloak medium, a negative uniaxial medium, gives an
ellipse with the major axis along the $-k_z$ axis.  The phase
matching condition is represented by the line at $k_x = k_{ix}$ and
determines the transmitted wave vector $k_t$. Note that $k_t$
undergoes a refraction as if the medium is optically denser, but
since the medium is anisotropic, the normal of the $k$-surface
determines the direction of power propagation, which is denoted by
the arrow, $<S_t>$.  Thus, the transmitted ray is refracted at an
angle $\theta_t > \theta_i$, while maintaining the phase in the $z$
direction, just as desired. Therefore, the original reflected beam
in the region of interest is recovered, producing a perfect cloak.

We can use the exact solution of Maxwell's equations for an incident
Gaussian beam~\cite{biwu} to illustrate this point. Figure
\ref{planecloak}(b) and (d) show a Gaussian beam incident at
$45^{\circ}$ to the PEC ground plane and the same beam incident onto
the cloak, respectively, corresponding to the ray tracing diagrams
in Fig.~\ref{planecloak}(a) and (c). The instantaneous $E$ field is
plotted. The parameters of the Gaussian beam are so chosen that the
phase and energy propagation can be clearly seen. It can be seen
that the phase and energy propagation directions are refracted at
different angles in Fig.~\ref{planecloak}(d), resulting in the
fields being identical to Fig.~\ref{planecloak}(b) above $z=d$.

It is interesting to investigate what will happen if we use an
isotropic medium instead. The desired anisotropic medium has two
principle refractive indices $n_L$ and $n_T$~\cite{li_carpet}, where
in our case $n_L =1$ and $n_T=1/(1-\delta)$. The refractive index of
the candidate isotropic medium is chosen to be the geometrical
average $n=\sqrt{n_L n_T}$~\cite{li_carpet} which means an
infinitesimal area should be preserved in the approximate
implementation. Therefore, the radius of the circular $k$ surface of
this isotropic medium is between the shorter and longer axes of the
elliptical $k$ surface of the anisotropic medium, as the dotted
circle shows in Fig.~\ref{phasematching}. With the same incident
beam, the phase matching condition determines the refracted
$\tilde{k}_t$. Because of isotropy, the transmitted Poynting power
$<\tilde{S}_t>$ is in the same direction as $\tilde{k}_t$. It can be
seen that $\tilde{k}_t$ is very close to $k_t$, meaning that the
emitting ray will possess a phase close to the case of a perfect
cloak, while $<\tilde{S}_t>$ deviates from $<S_t>$ with a large
angle $\Delta \theta_t$, meaning that the emitting ray will be
laterally shifted by a perceptible amount.
Figure~\ref{planecloak}(e) and (f) clearly show the ray tracing
diagram and the Gaussian beam incident onto this isotropic medium.
It can be seen that although the emitting ray as a whole has been
shifted by a large amount, the phase of this emitting ray does not
change much.

We have thus demonstrated the fundamental role of anisotropy using a
relatively large anisotropy of 2. We now examine the effects of
small anisotropy. In the case of Fig.~\ref{planecloak}(c), if we let
$\delta$ be small, the anisotropy $\alpha$ will be small. In this
case the anisotropy $\alpha$ is homogeneous and to further reduce
$\alpha$ at one location will cause increase of $\alpha$ at another
location. In other words, the anisotropy is already minimal.
According to Ref.~\cite{li_carpet}, if $\alpha$ is very small, we
can drop it and replace the anisotropic material by isotropic
material with refractive index $n=\sqrt{n_L n_T}$. For a ray
incident on the perfect cloak with incident angle $\theta_i$, the
distance between the incident point and the emitting point, i.e. the
segment AC in Fig.~\ref{planecloak}(c) is expected to be $s=2d\tan
\theta_i$, since it is equivalent to a virtual space with height of
$d$. However, in the isotropic case, this distance is
$\tilde{s}=2(1-\delta) d \frac{\sin
\theta_i/n}{\sqrt{1-\sin^2\theta_i /n^2}}< 2d(1-\delta) \tan
\theta_i$. Therefore the reflected ray is laterally shifted toward
the incident point by $s-\tilde{s} > 2\delta d \tan \theta_i =
O(\delta d)$, which is proportional to the height $\delta d$ of the
object to be hidden, even if the thickness of the cloak $d$ is
greatly increased.

\begin{figure}
\includegraphics[width=1\columnwidth,draft=false]{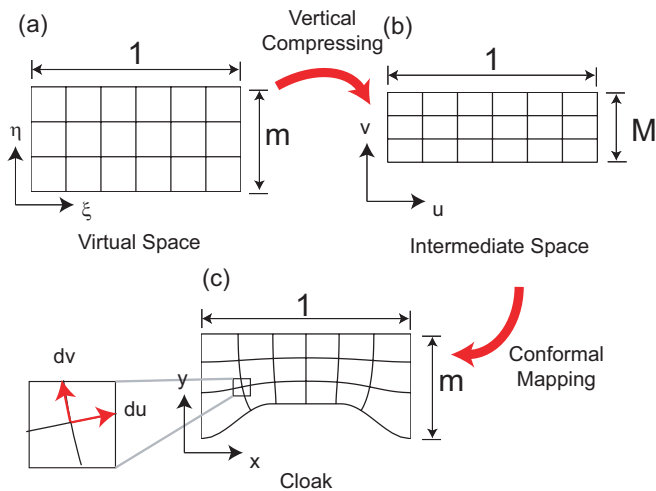}
  \caption{\label{mapping_relation} (Color online) Procedure of quasiconformal mapping. (a) A rectangle of $1 \times m$ in virtual free space. (b) Compressed rectangle of $1\times M$ in the intermediate space. (c) An anisotropic cloak created from quasiconformal mapping.}
\end{figure}

We now consider a curved bump similar to the case in
Ref.~\cite{li_carpet}, as shown in Fig.~\ref{raytracing}. The cloak
is created from a quasiconformal mapping, which omits an anisotropy
ratio of 1.042~\cite{li_carpet}. The ray tracing exercise in
Fig.~\ref{raytracing} shows that there is indeed a lateral shift of
the reflected ray comparable to the maximum height of the bump. Note
that the incident point of the ray on the surface of the bump is not
on the exact peak of the bump, but a little bit in front of the
peak. This is also caused by the omission of anisotropy. However, if
we examine the phase of the ray, due to a similar reason of two
close wave vectors $k_t$ and $\tilde{k}_t$ as shown in
Fig.~\ref{phasematching} (in the present case these two vectors are
much closer due to smaller anisotropy), the phase propagation can be
preserved to a large extent, whose distortion is difficult to
perceive.

Now we discuss the origin of this shift in the quasiconformal
mapping-based isotropic ground-plane cloak. In
Fig.~\ref{mapping_relation}, we first show the procedure of such
quasiconformal mapping. The virtual free space that appears as a
rectangle with dimensions $1\times m$
(Fig.~\ref{mapping_relation}(a)) needs to be compressed vertically
to fit the rectangle of dimensions $1 \times M$ in the intermediate
space (Fig.~\ref{mapping_relation}(b)), where $M$ is the conformal
module of the profile of the cloak with one curved
boundary~\cite{li_carpet} (Fig.~\ref{mapping_relation}(c)). From the
intermediate space to the cloak is a typical conformal mapping with
slipping boundary conditions. The anisotropy of $\alpha = m/M$ in
the anisotropic cloak~\cite{li_carpet} is inherited from the
homogeneous anisotropy of $\alpha = m/M$ in the intermediate space
in Fig.~~\ref{planecloak}(b) that is similar to the case we
discussed in Fig.~\ref{planecloak}(c). The assumption of the
quasiconformally mapped isotropic cloak is that the minimized
anisotropy can be dropped by letting $\alpha =1$ and taking the
refractive index $n = \sqrt{n_L n_T}$. We now study the consequence
of this operation and consider the inverse procedure of
Fig.~\ref{mapping_relation}. As shown in
Fig.~\ref{mapping_relation}(c), every mesh point of the anisotropic
cloak has two orthogonal directions, corresponding to principal
directions in the compressed rectangle
(Fig.~\ref{mapping_relation}(b)). In the ground-plane cloak the
anisotropy is $\frac{m}{M}$, because in
Fig.~\ref{mapping_relation}(c), the metric in $dv$ direction
$\sqrt{g_{vv}}$ and the metric in $du$ direction $\sqrt{g_{uu}}$
satisfy the relation $\sqrt{g_{vv}} = \frac{m}{M}\sqrt{g_{uu}}$.
When we force the two different principle refractive indexes $n_L$
and $n_T$ to be $n=\sqrt{n_L n_T}$, it is equivalent to decreasing
the metric along $dv$ direction and increasing the metric along $du$
direction by the same factor of $\sqrt{\alpha}$. With the inverse
conformal mapping, in the intermediate space, the metric along $du$
direction is increased from $1$ to $\sqrt{\alpha}$, while the metric
along $dv$ direction is decreased from $\alpha$ to $\sqrt{\alpha}$,
meaning now the intermediate space is an isotropic space. What
happens in the virtual free space is that its height becomes thinner
and its width becomes wider. It should be pointed out that due to
the slipping boundary condition in the quasiconformal mapping, the
interface of the top of the cloak is not perfectly matched to the
free space above the cloak. Therefore there is some additional
distortion caused by this mismatch. However, for the current case we
neglect this distortion.

Now we consider the virtual space of the example in
Fig.~\ref{raytracing}. Because of the smaller height, the distance
between the impinging point and the emitting point of the incident
and reflected rays in the virtual space (similar to $AC$ in
Fig.~\ref{planecloak}) is $2(h/\sqrt{\alpha})\tan{45^\circ}$. In
order to fit the wider width into the physical space, we need to
squeeze this segment $AC$ by $\sqrt{\alpha}$ to $2(h/\alpha)\tan
45^\circ$. We also need to consider that due to the slipping
boundary condition, the compression of mesh in the middle is denser
that other places. So the squeezed segment $AC$ will be a little
shorter and the final lateral shift caused by dropping anisotropy is
larger than $2h\tan 45^\circ - 2(h/\alpha)\tan 45^\circ \approx
0.12$ units. In our ray tracing exercise, the final lateral shift is
about 0.15 units. Note that this shift at $45^\circ$ incidence can
be reduced by increasing the width of the cloak. However, if the
cloak is very wide, the incident wave with the largest incident
angle can have a very long trajectory inside the cloak, which will
accumulate a perceptible lateral shift.

In conclusion, the invisibility performance of the quasiconformal
mapped isotropic ground-plane cloak is examined.  A lateral shift of
scattered waves, being comparable to the height of the hidden
object, will result when anisotropy is neglected. This is because
the corresponding virtual space is thinner and wider than the ideal
case. Ray tracing and theoretical analysis confirm our point on a
concrete example.

This work is sponsored by the Department of the Air Force under Air
Force Contract No.FA8721-05-C-0002.

\end{document}